\documentclass[12pt]{article}

\usepackage{graphicx}
\usepackage{amssymb}
\usepackage{amsmath}

\font\mybb=msbm10 at 10pt
\def\bb#1{\hbox{\mybb#1}}

\def\bC {\bb{C}}

\def\bfn{\mbox{\boldmath $\nabla$}}

\begin{document}

  \textheight=24cm
  \textwidth=16.5cm
  \topmargin=-1.5cm
  \oddsidemargin=-0.25cm

\begin{titlepage}

\begin{center}

\vskip 1.5cm

{\large \bf On non-relativistic 3D Spin-1 theories$^*$}

\vskip 1cm

{\bf  Eric A.~Bergshoeff$^1$, Jan Rosseel$^2$ and Paul K.~Townsend$^3$}

\vskip 25pt

{\em $^1$ \hskip -.1truecm Centre for Theoretical Physics,
University of Groningen, \\ Nijenborgh 4, 9747 AG Groningen, The
Netherlands \vskip 5pt }

\vskip .4truecm

{\em $^2$ \hskip -.1truecm Faculty of Physics, University of Vienna, Boltzmanngasse 5,\\  A-1090 Vienna, Austria \vskip 5pt }

\vskip .4truecm

{\em $^3$ \hskip -.1truecm  Department of Applied Mathematics and Theoretical Physics, Centre for \\ Mathematical Sciences, University of Cambridge,
Wilberforce Road, Cambridge, CB3 0WA, U.K.\vskip 5pt }

\vskip .4truecm

{emails: {\tt E.A.Bergshoeff@rug.nl}, {\tt rosseelj@gmail.com} \\ and {\tt p.k.townsend@damtp.cam.ac.uk}}
\end{center}

\vskip 1.5cm

\begin{center} {\bf ABSTRACT}\\[3ex]
\end{center}
We describe non-relativistic limits of the 3D Proca and $\sqrt{\rm Proca}$ theories that yield  spin-1 Schroedinger equations.
 Analogous results are found by generalized null reduction of the  4D Maxwell or complex self-dual Maxwell equations.
 We briefly discuss the extension to spin-2.

\vskip 4.3truecm

\noindent $^*$ {\small Contribution to the proceedings of the International Workshop {\it Supersymmetries and Quantum Symmetries SQS'2017}, Dubna, July 31 - August 5 2017.}

\end{titlepage}

\newpage

\section{Introduction}
Relativistic field theories of massless particles, such as Maxwell's electrodynamics or Einstein's General Relativity (GR), have non-relativistic limits in which these particles disappear,
in accordance with the instantaneous action at a distance of electro/magneto-statics or  Newtonian gravity. Put differently, they disappear because their velocity of propagation has become infinite. This suggests that  relativistic theories of {\it massive} particles, which move subluminally,  should  survive a non-relativistic limit, but this is not generally true. For example, one can take the speed of light to infinity in the
Klein-Gordon (KG) equation  if the particle's Compton wavelength is kept fixed, but the result is a Yukawa/Laplace equation that does {\it not} propagate any disturbance.

However, if the KG scalar field is {\it complex} then a different non-relativistic limit is possible, and this yields a Schroedinger equation for a massive particle of zero spin.
A similar limit is possible for relativistic  tensor field equations describing massive particles of non-zero integer spin, such as the Proca equations for spin-1 or the Fierz-Pauli (FP) equations for  spin-2,
but again only if the tensor field is complex. However, the 3D case (i.e. field theories in a spacetime of $2+1$ dimensions) is an exception to this rule.
As we have recently shown \cite{Bergshoeff:2017vjg}, the 3D FP equations have a novel non-relativistic limit to a planar spin-2 Schroedinger equation proposed previously in the context of the ``gapped'' spin-2 GMP mode of fractional Quantum Hall states\footnote{See \cite{Gromov:2017qeb} for a recent discussion  with references to the condensed matter literature.}.

The 3D case is also special in another respect: one can take the ``square-root'' of the Proca equation \cite{Townsend:1983xs} and of the spin-2 FP equation \cite{Aragone:1986hm}. These first-order equations, which  propagate  a single mode rather than a parity doublet, are equivalent to linearizations of parity-violating  ``topologically massive'' gauge theories, such as ``topologically massive gravity''  in the spin-2 case \cite{Deser:1981wh}; a  systematic derivation of such  equivalences may be found in \cite{Bergshoeff:2009tb}.  For such theories a non-relativistic limit to a planar Schroedinger equation requires  a complex field. Here we describe this limit for complex $\sqrt{\rm Proca}$ and show that it yields the same  planar spin-1 Schroedinger equation that one finds from an application of the ``novel non-relativistic limit'' of \cite{Bergshoeff:2017vjg} to real-field Proca. Moreover, we do this at the level of the actions, not just the equations.

It was also shown in \cite{Bergshoeff:2017vjg} that a generalized null reduction of the linearized 4D Einstein equations leads to the same  planar spin-2 Schroedinger equation as found from
the non-relativistic limit of the 3D real-field FP equations. This is  a further illustration of the long-established connection between the Galilei and Lorentz groups in $d$ and $d+1$ dimensions \cite{Gomis:1978mv}, and may  be compared with the derivation of 4D Newtonian gravity from 5D GR \cite{Duval:1984cj,Julia:1994bs}. Here we consider the same generalized null reduction for
the 4D Maxwell equations (real, complex or complex self-dual) in Bargmann-Wigner form.  We find a complete correspondence with the 3D non-relativistic limit results.

In our conclusions, we extract some general lessons and discuss briefly the extension of our results to spin-2 and beyond. For simplicity, we set
$\hbar=1$ throughout.

\section{The non-relativistic limit for 3D spin-1}

Non-relativistic limits are most simply investigated at the level of field equations but it is also of interest to know whether, and if so how, the limit
can be taken in the action. As the spin-1 cases considered here are relatively simple, we shall consider the non-relativistic limits of the complex
Proca and $\sqrt{\rm Proca}$ theories, in a 3D Minkowski vacuum,  at the level of the action. In both cases the Proca field is a 3-vector
$A_\mu$ ($\mu=0,1,2$) but an additional auxilary vector field will be needed for the $\sqrt{\rm Proca}$ case.

\subsection{Complex 3D Proca}

The Lagrangian density for a complex 3D Proca field $A_\mu$  ($\mu=0,1,2$) is
\begin{equation}\label{Procalag}
{\cal L} = -\frac{1}{4} \sqrt{-\det\eta} \, \eta^{\mu\rho}\eta^{\nu\sigma}  F^*_{\mu\nu} F_{\rho\sigma} - \frac{1}{2} (mc)^2 \sqrt{-\det\eta}\, \eta^{\mu\nu} A^*_\mu A_\nu\, ,
\end{equation}
where $c$ is the speed of light and $\eta_{\mu\nu} = {\rm diag.}(-c^2,1,1)$ is the 3D Minkowski metric.  After a time-space split $\mu=(0,i)$ ($i=1,2$),

\begin{equation}\label{cProca}
c^{-1} {\cal L} = -\frac{1}{4} F^\star_{ij}F_{ij} + \frac{1}{2c^2} F^\star_{0i}F_{0i} + \frac{m^2}{2} A^\star_0A_0 - \frac{(mc)^2}{2} A^\star_i A_i\,,
\end{equation}
where $F_{ij} = 2\partial_{[i }A_{j ]}$ and $F_{0i}=\dot A_i-\partial_iA_0$,  and there is an implied sum over repeated 2-space indices.
We now define new complex fields $(a_0,a_i)$ by setting
\begin{equation}
A_0 = e^{-imc^2t}a_0\, , \qquad A_i = e^{-imc^2t}a_i\, .
\end{equation}
After substitution into (\ref{cProca}), one finds a cancellation of terms on the right hand side of (\ref{cProca}) that diverge as $c\to\infty$. The subleading, and Galilean invariant, terms  are
\begin{equation}
{\cal L}_{\textrm{non-rel}} =  -\frac{1}{4} f^\star_{ij}f_{ij} +\frac{i}{2}ma_i^\star f_{0i} -\frac{i}{2}ma_if^\star_{0i} + \frac{m^2}{2} a_0^\star a_0\,,
\end{equation}
where $f_{ij} = 2\partial_{[i }a_{j]}$ and $f_{0i} = \dot a_i-\partial_ia_0$.  The $a_0$ field is auxiliary and can be eliminated;
omitting a total derivative, this yields
\begin{equation}
{\cal L}_{\textrm{non-rel}}  = \frac{1}{2}a_i^* \nabla^2 a_i + \frac{i}{2}m\left(a_i^* \dot a_i - a_i \dot a_i^*\right) \, , \qquad \nabla^2 = \partial_1^2 + \partial_2^2\, .
\end{equation}
The field equation is
\begin{equation}\label{preS}
2mi {\dot a}_i +\nabla^2 a_i=0\,,
\end{equation}
which  is an $SO(2)$-rotation doublet of Schroedinger equations.

In terms of the following helicity eigenfunctions
\begin{equation}
\Psi[1] = a_1 + ia_2\, , \qquad \Psi[-1] = a_1^* + i a_2^*\, ,
\end{equation}
the equations (\ref{preS}) become
\begin{equation}
i\dot \Psi[\pm1] = \mp \frac{1}{2m} \nabla^2 \Psi[\pm 1]\, .
\end{equation}
The two equations are exchanged by parity, but we may impose the self-duality constraint
\begin{equation}\label{duality}
a_i = \mp i\epsilon_{ij} a_j\, ,
\end{equation}
which implies that $\Psi[\mp1]=0$, thereby leaving a single parity-violating planar spin-1 Schroedinger equation for $\Psi[\pm1]$. Assuming
that $m$ is positive we must choose to retain the equation for $\Psi[1]$ in order to have a positive Hamiltonian.

\subsection{Complex 3D $\sqrt{\rm Proca}$}

We now turn to the complex $\sqrt{\textrm{Proca}}$ theory.  It turns out that to take the non-relativistic limit in the action we must include a complex auxiliary vector field $B_\mu$; the
Lagrangian density is
\begin{equation}
{\cal L}= \varepsilon^{\mu\nu\rho} A^*_\mu \partial_\nu A_\rho -(mc) \sqrt{-\det\eta}\, \eta^{\mu\nu}(A^*_\mu A_\nu - 2B^*_\mu B_\nu)
\end{equation}
or, after a time-space split,
\begin{eqnarray}\label{sqrtproca}
{\cal L}_{\textrm{rel}} &=& -\epsilon^{ij}A_i^\star \dot A_j  - (mc^2)(A^*_i A_i  -  2B^*_i B_i)\nonumber \\
&& \quad +\,   \epsilon^{ij}(A_0^\star \partial_i A_j + c.c.) + m (A_0^*A_0 - 2B_0^*B_0)\, .
\end{eqnarray}
We can eliminate the auxiliary fields $A_0$ and $B_0$ to get
\begin{equation}
{\cal L}=  -\epsilon^{ij}A_i^\star \dot A_j  - (mc^2)(A^*_i A_i  -  2B^*_i B_i) - \frac{1}{m} \left|\epsilon^{ij}\partial_i A_j\right|^2\, .
\end{equation}

We now define new  complex fields $(a_i,b_i)$  by
\begin{equation}\label{redef2}
A_i = e^{-imc^2t}a_i  \, , \qquad  B_i = e^{-imc^2t}\left([{\cal P}a]_i  + \frac{1}{2mc^2} b_i\right)\, ,
\end{equation}
where ${\cal P}$ is a complex projector matrix:
\begin{equation}
{\cal P}_{ij} = \frac{1}{2} \left(\delta_{ij} -  i\epsilon_{ij}\right)\, , \qquad {\cal P}^2={\cal P}\, , \quad {\cal P}\bar{\cal P} = \bar{\cal P}{\cal P} =0\, .
\end{equation}
Substitution yields a Lagrangian in which the terms proportional to $c^2$ have cancelled.  Taking the $c\to\infty$ limit, we arrive at
\begin{equation}\label{abeqs}
{\cal L}_{\rm non-rel} = - \epsilon^{ij} a^*_i \dot a_j  - \frac{1}{m} \left|\epsilon^{ij}\partial_i a_j\right|^2 +  \left( b^*_i [Pa]_i + c.c.\right) \, .
\end{equation}
The projection  $\bar{\cal P}b$ of $b$ no longer appears because it  drops out in the $c\to\infty$ limit; the other projection  of $b$
is a Lagrange multiplier for the constraint ${\cal P}a=0$, which is equivalent to
\begin{equation}\label{idents}
a_i = \left[\bar{\cal P}a\right]_i\, .
\end{equation}
This implies that  $\epsilon^{ij}\partial_i a_j  = -i\partial_i a_i$, which can be used to show that
\begin{equation}\label{hodge}
\left[\bar{\cal P} \partial\right]_i (\epsilon^{kl}\partial_k a_l)) = -\frac{i}{2} \nabla^2 a_i \, .
\end{equation}

The field equation found from variation of $a_i$ in (\ref{abeqs}) is
\begin{equation}\label{aeqs}
\left[{\cal P}b\right]_i = \epsilon^{ij}\dot a_j + m^{-1} \epsilon^{ij}\partial_j \left( \epsilon^{kl}\partial_k a_l\right) \, .
\end{equation}
This determines ${\cal P}b$, which is therefore auxiliary, but it also implies that
\begin{equation}\label{aeq}
m\left[\bar {\cal P}\dot a\right]_i + \left[\bar{\cal P} \partial\right]_i (\epsilon^{kl}\partial_k a_\ell) =0\, .
\end{equation}
Using (\ref{idents}) and (\ref{hodge}), we may rewrite this as
\begin{equation}
2mi \dot a_i + \nabla^2 a_i =0\, .
\end{equation}
This is  (\ref{preS}), but here $a$ is constrained to satisfy  $a_i= i\epsilon_{ij}a_j$; as we saw, this is equivalent to $\Psi[-1]=0$,
which leaves a single parity-violating Schroedinger equation for $\Psi[1]$, with positive Hamiltonian if $m$ is positive.
We saw previously
that this truncation could be imposed  `by hand' but here it is the result of a  field equation.

If the sign of $m$  is changed then we can maintain the positivity of the Hamiltonian if we also replace ${\cal P}$ by $\bar {\cal P}$ in (\ref{redef2}), but
then we get a Schroedinger equation for $\Psi[-1]$ in place of $\Psi[1]$.

\subsection{Real Proca}

We are now going to see how the above result for the non-relativistic limit of  $\sqrt{\rm Proca}$ can also be found by taking a novel non-relativistic
limit of {\it real} 3D Proca. To this end, we return to the Proca Lagrangian (\ref{Procalag}) but now for a {\it real} vector field $A_\mu=(A_0,{\bf A})$.
A time-space decomposition then yields
\begin{equation}
{\cal L}= \frac{1}{2c^2} |\dot{\bf A}|^2 - \frac{1}{2} {\bf A} \cdot \Delta {\bf A} +
\frac{1}{2} (\bfn\cdot {\bf A})^2 + \frac{1}{2c^2} A_0 \Delta A_0 + \frac{1}{c^2} A_0 (\bfn\cdot \dot{\bf A})\, ,
\end{equation}
where
\begin{equation}
\Delta = - \nabla^2 + (mc)^2 >0\, .
\end{equation}
Elimination of the auxiliary field $A_0$ yields the following Lagrangian
\begin{equation}
{\cal L} = \frac{1}{2c^2} \dot A_i K_{ij} \dot A_j - \frac{1}{2} A_i \Delta K_{ij} A_j
\end{equation}
where $A_i$ ($i=1,2$) are the components of ${\bf A}$ and
\begin{equation}
K_{ij} = \delta_{ij} + \Delta^{-1} \partial_i \partial_j\, .
\end{equation}
These are the entries of a matrix $K$, with inverse
\begin{equation}
K^{-1}_{ij} = \delta_{ij} - (mc)^{-2} \partial_i\partial_j\, .
\end{equation}

Next,  we set
\begin{equation}\label{Bdef}
A_i = [K^{-\tfrac{1}{2}}]_{ij} B_j
\end{equation}
for a new 2-vector field ${\bf B}$.  In terms of ${\bf B}$, the Lagrangian density is
\begin{equation}
{\cal L} = \frac{1}{2c^2} |\dot{\bf B}|^2 - \frac{1}{2} {\bf B} \cdot \Delta {\bf B} \, .
\end{equation}
In terms of the complex field $B= (B_1+ iB_2)/\sqrt{2}$, this is
\begin{equation}\label{spin1}
{\cal L} = \frac{1}{c^2} |\dot B|^2 + \bar B  \nabla^2 B - (mc)^2|B|^2\, .
\end{equation}
If we now set
\begin{equation}
B= e^{-imc^2 t} \Psi[1]\, ,
\end{equation}
for new complex variable $\Psi[1]$, we may take the $c\to\infty$ limit to arrive at the Galilean invariant Lagrangian density
\begin{equation}
{\cal L}_{NR}  = 2im\bar\Psi[1] \dot\Psi[1] + \bar\Psi[1] \nabla^2  \Psi[1]\, .
\end{equation}
The field equation is
\begin{equation}
- \frac{1}{2m} \nabla^2\Psi[1] =  i \dot\Psi[1]\, ,
\end{equation}
which is the planar spin-1 Schroedinger equation.

\section{Spin-1 Schroedinger from 4D Maxwell}

We will now explain how all the spin-1 planar Schroedinger equations found above from non-relativistic limits of real or complex Proca, and complex $\sqrt{\rm Proca}$,  can also be found from
a generalized null reduction of the real or complex 4D Maxwell equations, and the complex self-dual Maxwell equations. We work directly with equations, rather than the action, and we
start from the Bargmann-Wigner (BW) form of Maxwell's equations for a complex symmetric $Sl(2;\bC)$ bi-spinor $F_{\alpha\beta}$ ($\alpha,\beta=1,2$). In Fourier space, the BW equations are
\begin{equation}\label{BWeq}
p^{\alpha\dot\alpha} F_{\alpha\beta} =0\,,
\end{equation}
where, in the $Sl(2;\bC)$ spinor conventions spelled out in \cite{Mezincescu:2015apa},
\begin{equation}
p^{\alpha\dot\alpha}  = \left(\begin{array}{cc} - \sqrt{2} p_- & - p \\ - \bar p & \sqrt{2} p_+ \end{array} \right)\, , \qquad p=p_1+ip_2\, .
\end{equation}

Inspired by the Scherk-Schwarz dimensional reduction \cite{Scherk:1979zr}, we now effect a generalized null reduction by choosing\footnote{The choice $m=0$ corresponds to a standard null reduction, equivalent to supposing that all fields are independent of $x^-$, and does not lead to a Schroedinger equation. }
\begin{equation}\label{nullred}
p_-  = m
\end{equation}
for mass $m$. This choice is consistent with the fact that $p_-$ is the Fourier-dual of the complex differential operator $-i\partial_-$ because $F_{\alpha\beta}$ is complex, and it  gives us
\begin{equation}
p^{\alpha\dot\alpha}  =  \left(\begin{array}{cc}  -\sqrt{2}m  & - p \\ - \bar p & -\sqrt{2} E \end{array} \right)\, , \qquad E= -p_+ \, .
\end{equation}
Using this in the BW equations, we have
\begin{equation}
\left(\begin{array}{cc}  -\sqrt{2}m  & - p \\ - \bar p & -\sqrt{2} E \end{array} \right)\left(\begin{array}{c} F_{1\beta} \\ F_{2\beta} \end{array}\right) =0\, ,
\end{equation}
which is equivalent to
\begin{equation}
\sqrt{2}\, m F_{1\beta} = -pF_{2\beta} \, , \qquad \left[2mE - |{\bf p}|^2\right]F_{2\beta} =0\, .
\end{equation}
In other words, $F_{1\beta}$ is auxiliary and $F_{2\beta}$ satisfies a Schroedinger equation. Moreover,  because  $F_{\alpha\beta}=F_{\beta\alpha}$ the same applies to $F_{\alpha 1}$ and $F_{\alpha 2}$, so the only component of $F_{\alpha\beta}$ that is not auxiliary is $F_{22}$, and this satisfies
\begin{equation}\label{F22}
\left[2mE - |{\bf p}|^2\right]F_{22}=0\, ,
\end{equation}
which is a Schroedinger equation for a single complex wavefunction.

Taking the complex conjugate of this Schroedinger equation we find that
\begin{equation}\label{F22}
\left[2mE - |{\bf p}|^2\right]\bar F_{\dot 2\dot 2}=0\, .
\end{equation}
If we had started from the Maxwell equations for a {\it complex} vector potential then $F_{22}$ and $\bar F_{\dot 2\dot 2}$ would be {\it independent} complex wavefunctions
rather than complex conjugates of each other, and we would recover the parity-doublet of Schroedinger equations that we found from the non-relativistic limit of the complex 3D Proca equations.
However, in this case it is consistent to impose  $F_{\dot\alpha\dot\beta}=0$, {\it without this implying $F_{\alpha\beta}=0$}. This is equivalent to imposing a self-duality condition on the
complex Maxwell field-strength 2-form. This  effects the same truncation to the parity-violating spin-1 Schroedinger equation that we found from the non-relativistic limit of
the complex $\sqrt{\rm Proca}$ equations.

\section{Conclusions}

We have investigated the non-relativistic limits of free 3D field theories for spin-1 particles of non-zero mass $m$;  specifically, 3D Proca and $\sqrt{\rm Proca}$ and their complexifications.
A common feature is that the initial relativistic 3D theory must propagate {\it pairs} of modes of equal mass, since two are needed for every complex wavefunction satisfying a
Schroedinger equation in the limit. This condition is satisfied automatically for a complex vector field, and for the 3D Proca with real vector field, but not for $\sqrt{\rm Proca}$
with real vector field; in this last case the only non-relativistic limit is to equations that do not propagate any non-relativistic particle.

As the Proca theory preserves parity, one might expect its non-relativistic limit to also preserve parity and this is true if one starts from {\it complex}  3D Proca; its non-relativistic
limit is a Schroedinger equation for two complex wavefunctions transforming under rotations as an $SO(2)$ doublet. This system of equations is equivalent to equations for
a parity-doublet of complex helicity eigenstate wavefunctions $\Psi[\pm1]$  with opposite sign Hamiltonians, and the non-relativistic limit  of the parity-violating $\sqrt{\rm Proca}$
theory yields just one of these Schroedinger equations, which one depending on the sign of $m$.

In general, there is no non-relativistic limit of the {\it real} Proca theory to a Schroedinger equation preserving rotational invariance but the 3D case is an exception to this rule.
In this case there is such a limit, as we showed for the spin-2 FP equations in \cite{Bergshoeff:2017vjg}. Here, for the spin-1 case, we have shown how this novel non-relativistic limit may be taken in the
action (and not merely for  the field equations).  Somewhat surprisingly, it leads to the same parity-violating single Schroedinger equation that one gets from complex
$\sqrt{\rm Proca}$.

We have further demonstrated a correspondence between these results and those obtained from a generalized null reduction \cite{Bergshoeff:2017vjg} of 4D field theories
for spin-1 particles of zero mass. Specifically, a generalized null reduction of the Bargmann-Wigner equations equivalent to complex 4D Maxwell, complex self-dual Maxwell  and real 4D Maxwell
leads to the same spin-1 planar Schroedinger equations as 4D complex Proca, complex $\sqrt{\rm Proca}$ and real Proca, respectively.  Furthermore, there is a straightforward generalization to
any integer spin; for example,  the linearized 4D Einstein equations are equivalent to the spin-2  Bargmann-Wigner equations
\begin{equation}
p^{\alpha\dot\alpha} R_{\alpha\beta\gamma\delta} =0\, ,
\end{equation}
where $R_{\alpha\beta\gamma\delta}$ is totally symmetric in its four $Sl(2;\bC)$ indices.   Following the example of the spin 1 case we deduce that only $R_{2222}$ is independent, and that it satisfies
\begin{equation}
\left[2mE - |{\bf p}|^2\right]R_{2222}=0\, .
\end{equation}
This is equivalent to the generalized null-reduction of  \cite{Bergshoeff:2017vjg}.  Complexifying the linearized Einstein equations leads to independent equations for
$R_{2222}$ and $\bar R_{\dot 2\dot2\dot2\dot2}$, and the analog of Maxwell self-duality in this case is then $\bar R_{\dot 2\dot2\dot2\dot2}=0$.

Our results for non-relativistic limits of massive 3D spin-1 theories can also be generalized to any integer spin.  For example, the $\sqrt{\rm FP}$ action has a non-symmetric tensor field \cite{Aragone:1986hm}. After the addition of an FP-type mass term for an auxiliary non-symmetric tensor field, redefinitions similar to those of (\ref{redef2}) allow a non-relativistic limit to be taken. This
yields the {\it same} parity-violating planar spin-2 Schroedinger equation as that found in \cite{Bergshoeff:2017vjg} from a  ``novel non-relativistic limit'' of the real spin-2 FP equations.  However,
as mentioned there, the need for a complex field  in the $\sqrt{\rm FP}$  case complicates the issue of interactions because a metric perturbation is naturally real. This problem does not arise
for the real 3D FP equations because these result from linearization of the equations of ``New Massive Gravity'' \cite{Bergshoeff:2009hq}.

Finally, we should mention that non-relativistic limits of the Jackiw-Nair equations for massive particles of any spin have been proposed in \cite{Horvathy:2010vm}; the limits considered
there do not  include the novel real-field limit of \cite{Bergshoeff:2017vjg} that we have explained here for spin-1, but there is presumably an  overlap with the non-relativistic limit of the complex
$\sqrt{\rm Proca}$ case considered here.

\section*{Acknowledgements}

E.B. thanks the organizers of the SQS'2017 workshop for providing a stimulating atmosphere and
offering  a diverse scientific programme.
The work of PKT is partially supported by the STFC consolidated grant ST/P000681/1.

\end{document}